

\input eplain

\newcount\fignumber
\def\figdef#1{\global\advance\fignumber by 1 \definexref{#1}{\number\fignumber}{figure}\ref{#1}}
\def\figdefn#1{\global\advance\fignumber by 1 \definexref{#1}{\number\fignumber}{figure}}
\let\figref=\ref

\newcount\tabnumber
\def\tabdef#1{\global\advance\tabnumber by 1 \definexref{#1}{\number\tabnumber}{table}\ref{#1}}
\def\tabdefn#1{\global\advance\tabnumber by 1 \definexref{#1}{\number\tabnumber}{table}}

%
\ifx\pdfoutput\undefined
\input epsf

\def\figscale#1#2{\epsfxsize=#2\epsfbox{#1.eps}}
%
\else

\def\figscale#1#2{\pdfximage width#2 {#1.pdf}\pdfrefximage\pdflastximage}
\fi


\newcount\scount \scount=0



\makeatletter
\def\section#1\par{
  \vskip\z@ plus.3\vsize\penalty-250
  \vskip\z@ plus-.3\vsize\bigskip\vskip\parskip
  \global\advance\scount by1
  \writenumberedtocentry{section}#1{}
  \definexref#1{\the\scount}{section}
  \message{#1}
  \noindent\the\scount.\quad{\bf #1}\nobreak\smallskip\noindent}
\makeatother

\centerline{\bf{The Lamb-Oseen Vortex and Paint Marbling}}
\centerline{Aubrey G. Jaffer}
\centerline{agj@alum.mit.edu}

\beginsection{Abstract}

{\narrower

  The displacement pattern arising from the decay of a two-dimensional
  Lamb-Oseen vortex in a Newtonian fluid can be closely modeled by the
  closed-form expression presented here.  This formula enables
  Lamb-Oseen vortex simulation orders of magnitude faster than can be
  accomplished using finite-element methods, and without the
  accumulation of errors.

  ``French Curl'' marbling patterns look as though they are created by
  a vortex.  Analysis and simulation of a nineteenth century example
  of French Curl finds that the pattern was created without a vortex.

  True vortexes are rarely seen in paint marbling because, in order to
  reach Reynolds numbers larger than 90, viscosity of the fluid bath
  must be much lower than is customarily used.

  \par}

\beginsection{Keywords}

{\narrower Lamb-Oseen Vortex; paint marbling; Reynolds number; fuild mechanics\par}

\beginsection{Table of Contents}

\readtocfile

\section{Introduction}

 Marbling originated in Asia as a decorative art more than 800 years
 ago and has continued to evolve and thrive in parts of Asia and the
 Middle East.  Marbling spread to Europe in the 1500s where it was
 employed for endpapers and book covers.  Although Western interest
 declined with the advent of mechanized bookbinding, marbling has been
 enjoying a revival.  It can even be seen in the heart-shaped designs
 drawn in latte foam.

 Mathematically, it turns out that many common marbling techniques can
 be simply modeled by closed-form homeomorphisms\numberedfootnote{A
 homeomorphism is a continuous function between topological spaces (in
 this case between a topological space and itself) that has a
 continuous inverse function.}.  A paper illustrating this approach is
 {\it Mathematical Marbling}\cite{10.1109/MCG.2011.51}.  The fluid
 mechanics of marbling are explored in {\it Oseen Flow in Paint
 Marbling}\cite{2017arXiv170202106J}.

 The previous work modeled creeping flows, ignoring momentum.  The
 Lamb-Oseen vortex formula models the decay of the vortex over time,
 which includes momentum.  Surprisingly, the Lamb-Oseen vortex is
 reversible due to its laminar nature.

\section{Lamb-Oseen Vortex}

 In two dimensions, the Lamb-Oseen vortex is a flow arising from an
 impulse of circulation at the center point.  Its vorticity is nonzero
 only at the center point.  The radial velocity is zero everywhere.
 The fluid moves in concentric shells around the center.  The
 circumferential velocity is a continuous function of time $t$ and
 radius $r$ except at $r=0$.  Thus it is a laminar flow.  It is also
 reversible; a clockwise impulse followed by a counterclockwise
 impulse of the same magnitude eventually returns to the original
 position.



 Tryggeson\cite{tryggeson2007analytical} and
 Saffman\cite{saffman1995vortex} give the Lamb-Oseen formula for
 circumferential velocity of the irrotational vortex created by an
 impulse of circulation $\Gamma$ (in ${\rm m}^2/{\rm s}$) at its
 center:

$$u_\theta={\Gamma\over2\,\pi\,r}\left[1-\exp\left(-{r^2\over4\,\nu\,t}\right)\right]\eqdef{u_theta}$$



 In the log-log plot of \figref{lamb-oseen-time}, the dashed traces
 are the numerical integration of equation \eqref{u_theta} with
 respect to time $t$; circulation $\Gamma=10^{-3}~{\rm m^2/s}$.  The
 three longest-dashed traces (radii 1~cm, 10~cm, 1~m) have slopes near
 one, indicating linear growth with time; the other trace is at 1~mm
 radius.  The circumferential distance does not converge as
 $t\to\infty$.  But a marbling tank is not left for hours to settle.
 Keeping $t$ as a parameter to the formula, the only dimensionless
 combination of $r$, $t$, and $\nu$ from \eqref{u_theta} is
 $r^2\,\nu^{-1}\,t^{-1}$.  Combining this group with the constant 1
 using the $L^p$-norm with the unusual value of $p=-3/4$ in
 equation \eqref{fit} produces the values shown as squares
 in \figref{lamb-oseen-time}.

$$p_\theta={\Gamma\cdot t\over2\,\pi\,r}
 \cdot\left[1+\left({4\,\nu\,t\over2\,\pi\,r^2}\right)^{3/4}\right]^{-4/3}
 ={\Gamma\cdot t\over2\,\pi\,r}
 \cdot\left\|1,{2\,\pi\,r^2\over4\,\nu\,t}\right\|_{-3/4}
 \eqdef{fit}$$
where the $L^p$-norm is:
$$\left\|x,y\right\|_p=\left(|x|^p+|y|^p\right)^{1/p}$$

\vbox{\settabs 1\columns
\+\hfill\figscale{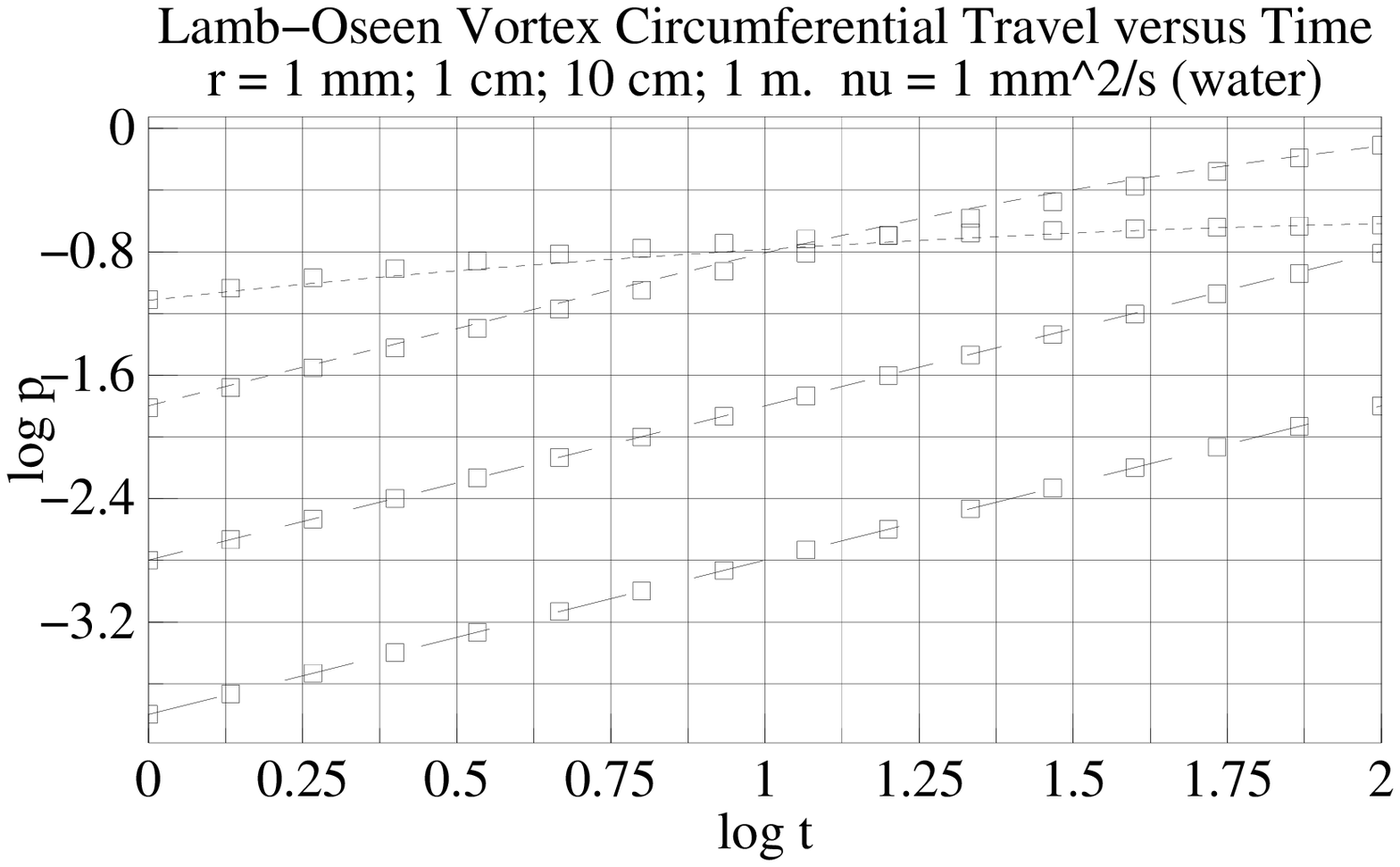}{250pt}\hfill&\cr
\+\hfill\figdef{lamb-oseen-time}\hfill&\cr
\+\hfill\figscale{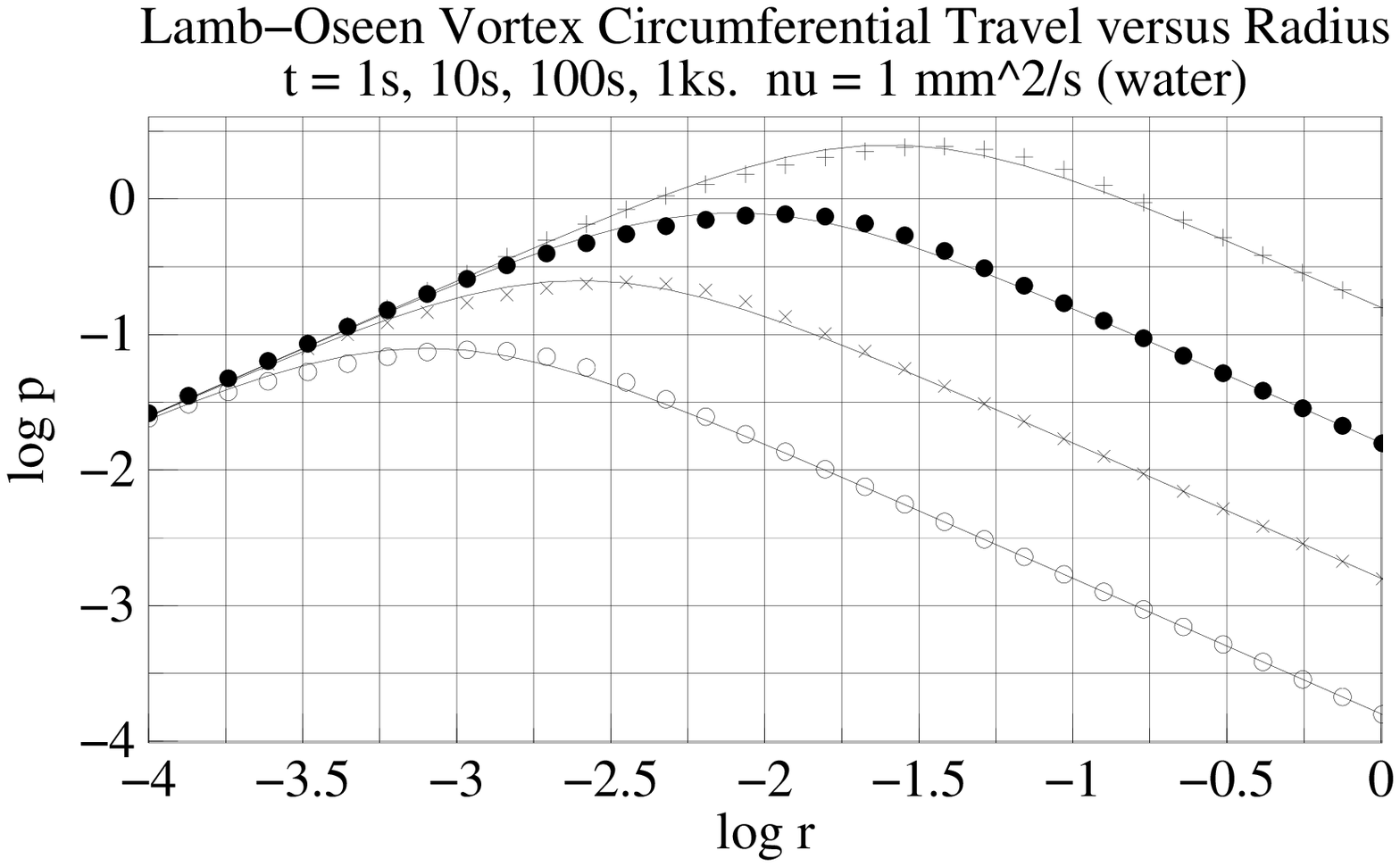}{250pt}\hfill&\cr
\+\hfill\figdef{lamb-oseen-rad}\hfill&\cr
}

 \figref{lamb-oseen-rad} shows the travel versus radius $r$ after 1~s
 (circles), 10~s (crosses), 100~s (dots), 1000~s (pluses); the lines
 are computed using formula \eqref{fit}.

 The derivative of formula \eqref{fit} with respect to $t$ should be
 similar to the formula \eqref{u_theta}:

$${d\,p_\theta\over d\,t}=
 {\Gamma\over2\,\pi\,r}\left[1+\left({4\,\nu\,t\over2\,\pi\,r^2}\right)^{3/4}\right]^{-7/3}
 ~{\buildrel?\over\approx}~
  {\Gamma\over2\,\pi\,r}\left[1-\exp\left(-{r^2\over4\,\nu\,t}\right)\right]=u_\theta$$

 The limit of both formulas as $t$ approaches 0 is
 $\Gamma/(2\,\pi\,r)$; the limit of both formulas as $t$ approaches
 $\infty$ is 0.  Letting $4\,\nu\,r^{-2}=1$ and $2\,\pi\,r/\Gamma=1$,
 $u_\theta$ is compared with the $L^{-3/4}$-norm in \figref{u-fit}.

\vbox{\settabs 1\columns
\+\hfill\figscale{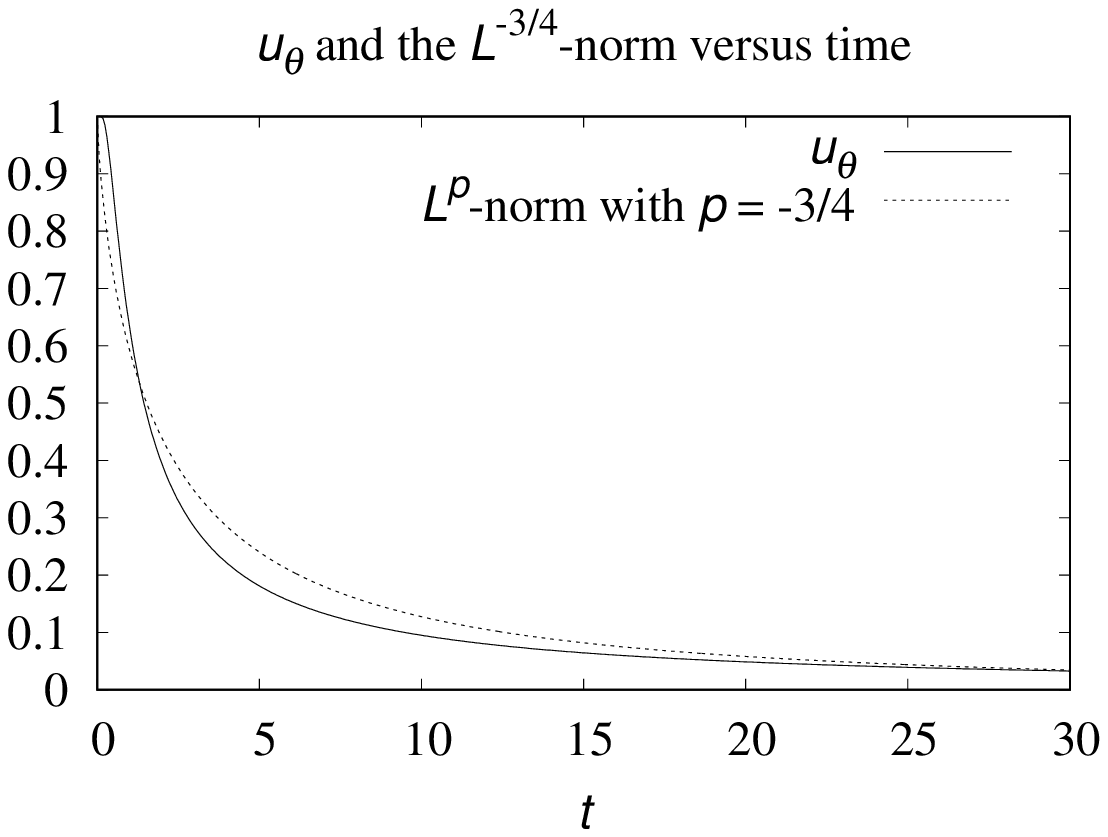}{250pt}\hfill&\cr
\+\hfill\figdef{u-fit}\hfill&\cr
\+\hfill\figscale{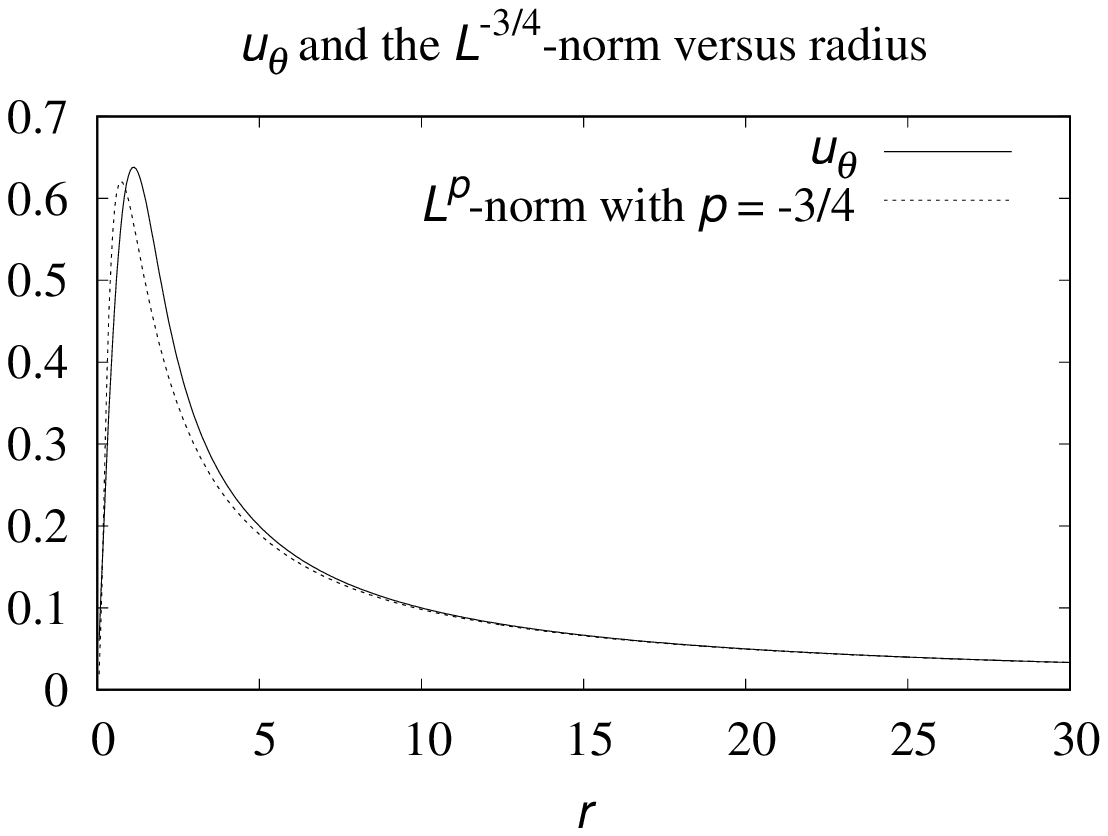}{250pt}\hfill&\cr
\+\hfill\figdef{u-fit-r}\hfill&\cr
}

 Letting $4\,\nu\,t=1$ and $2\,\pi/\Gamma=1$, $u_\theta$ is compared
 with the $L^{-3/4}$-norm in \figref{u-fit-r}.

\section{Angle}

 For the marbling transform it is more convenient to work with angle
 $a$ than circumferential distance $p_\theta$.  With the introduction
 of variable $\zeta={t/(2\,\pi\,r^2)}$, the formula simplifies:

$$a={p_\theta\over r}={\Gamma\,\zeta}
 \left[1+\left({4\,\nu\,\zeta}\right)^{3/4}\right]^{-4/3}
 ={\Gamma\,\zeta}\left\|1,{1\over4\,\nu\,\zeta}\right\|_{-3/4}
 =\Gamma\,\left\|\zeta,{1\over4\,\nu\,}\right\|_{-3/4}
 =\Gamma\,\left\|{t\over2\,\pi\,r^2},{1\over4\,\nu\,}\right\|_{-3/4}
 \eqdef{fita}$$

\vbox{\settabs 1\columns
\+\hfill\figscale{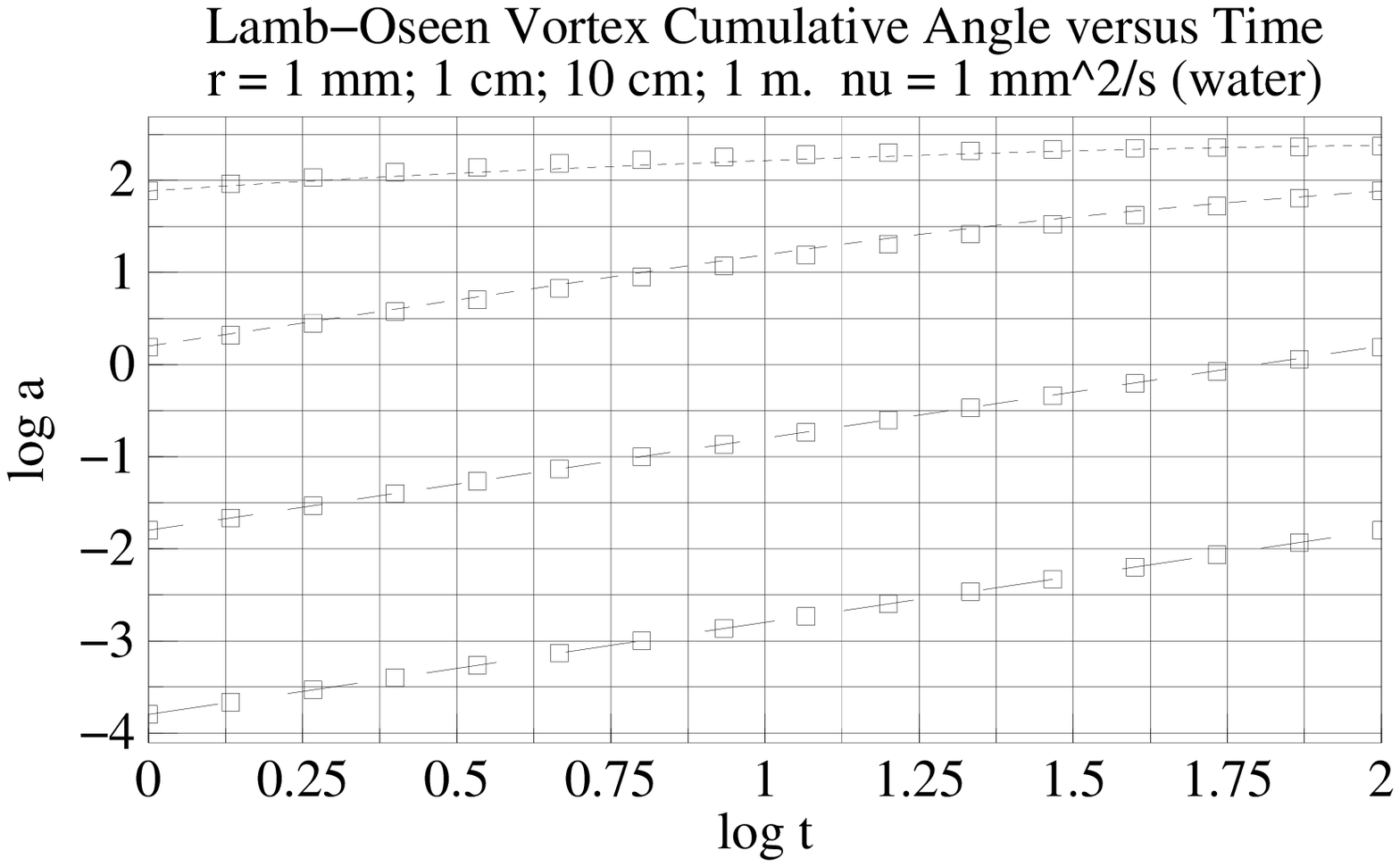}{250pt}\hfill&\cr
\+\hfill\figdef{lamb-oseen-time-ang}\hfill&\cr
\+\hfill\figscale{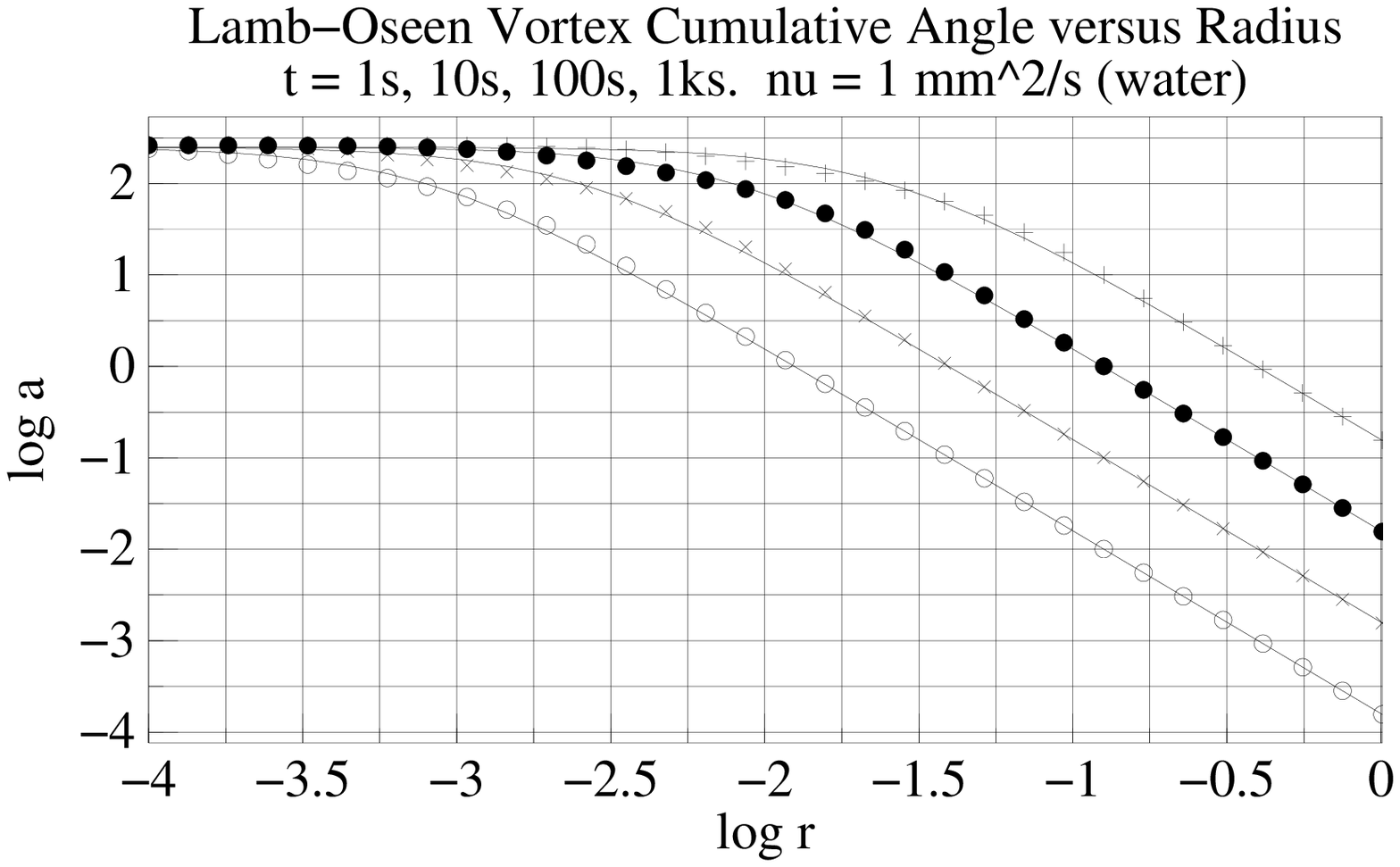}{250pt}\hfill&\cr
\+\hfill\figdef{lamb-oseen-rad-ang}\hfill&\cr
}

 As time increases, the traces in \figref{lamb-oseen-time-ang}
 eventually merge, resulting in the original pattern rigidly rotated.

 A vortex homeomorphism of circulation $\Gamma$ around $\vec C$ after
 time $t$ maps a point $\vec P$ to:

$$\vec C+\left[\vec P-\vec C\right]\cdot
   \pmatrix{\cos a&\sin a\cr-\sin a&\cos a\cr}
\qquad r=\left\|(\vec P-\vec C)\right\|$$

 \figref{irrot-0} shows a base pattern with no vortex.
 \figref{irrot-1} shows the pattern a short time after the impulse of circulation; the center of the vortex is rotated counterclockwise nearly $360^\circ$.
 After this the center rotates only a few degrees; the initial rotation propagates outward.
 \figref{irrot-2} shows the pattern after 10 times the duration as \figref{irrot-1}.
 \figref{irrot-3} shows the pattern after 100 times the duration as \figref{irrot-1}.
 After a very long time the pattern asymptotically returns to \figref{irrot-0}.
\bigskip
 On learning that the Lamb-Oseen vortex flow is reversible, Martin
 Jaffer immediately suggested that one could emulate sweeping freehand
 curves by applying the vortex homeomorphism, drawing a stylus on a
 straight path through it, then applying the reverse vortex
 homeomorphism, as shown in \figref{vortexline}.  Applying the vortex
 homeomorphism to a half-black, half-white base creates a design
 evocative of the yin-yang pattern in \figref{yinyang}.

\vbox{\settabs 2\columns\+
\hfill\figscale{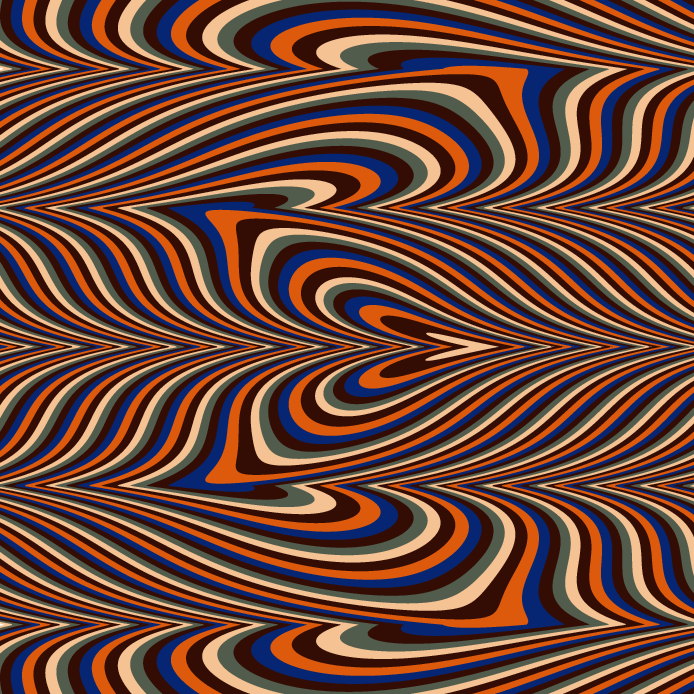}{200pt}\hfill&
\hfill\figscale{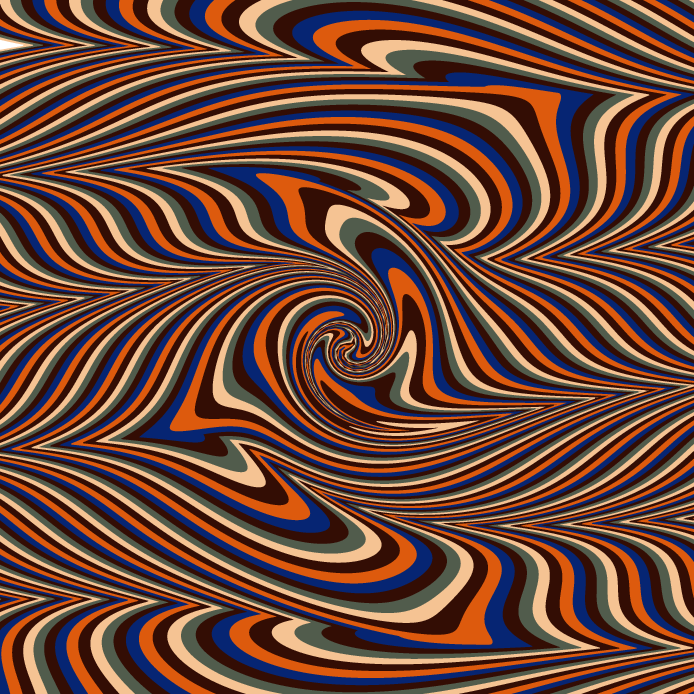}{200pt}\hfill&
\cr\+
\hfill\figdef{irrot-0}\hfill&
\hfill\figdef{irrot-1}\hfill&
\cr\+
\hfill\figscale{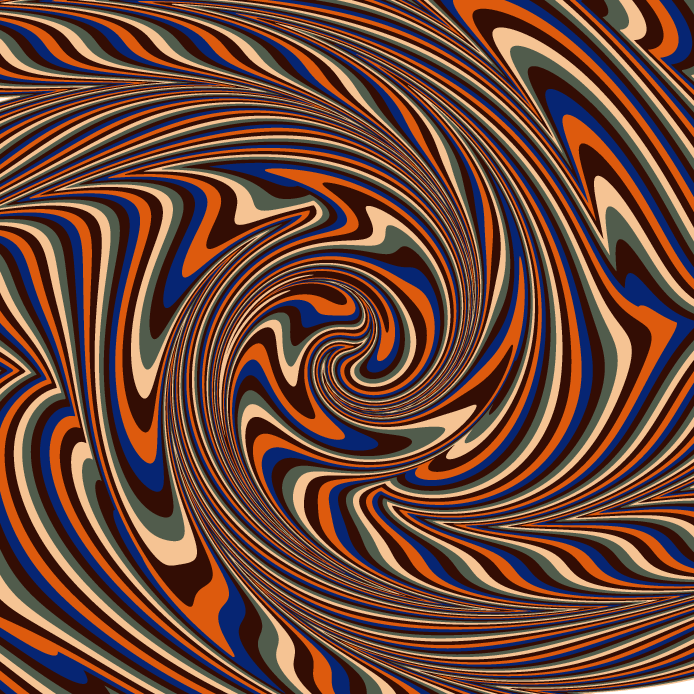}{200pt}\hfill&
\hfill\figscale{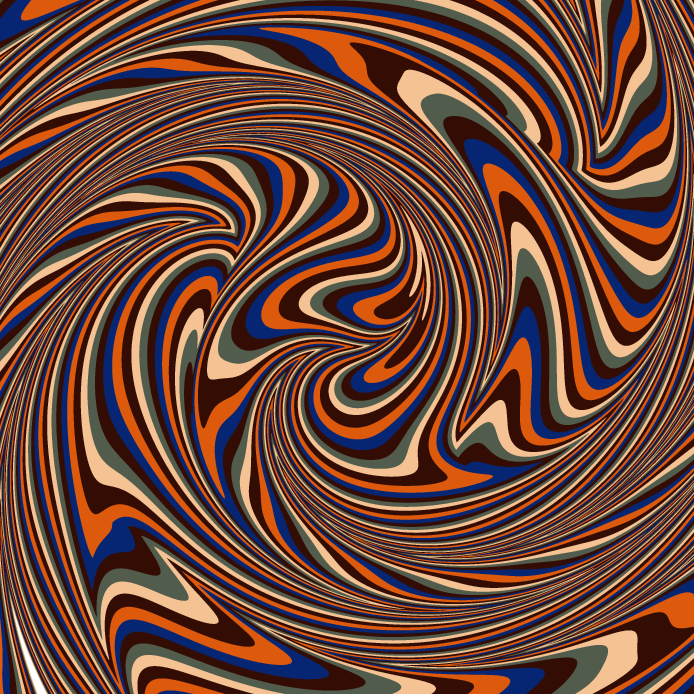}{200pt}\hfill&
\cr\+
\hfill\figdef{irrot-2}\hfill&
\hfill\figdef{irrot-3}\hfill&
\cr\+
\hfill\figscale{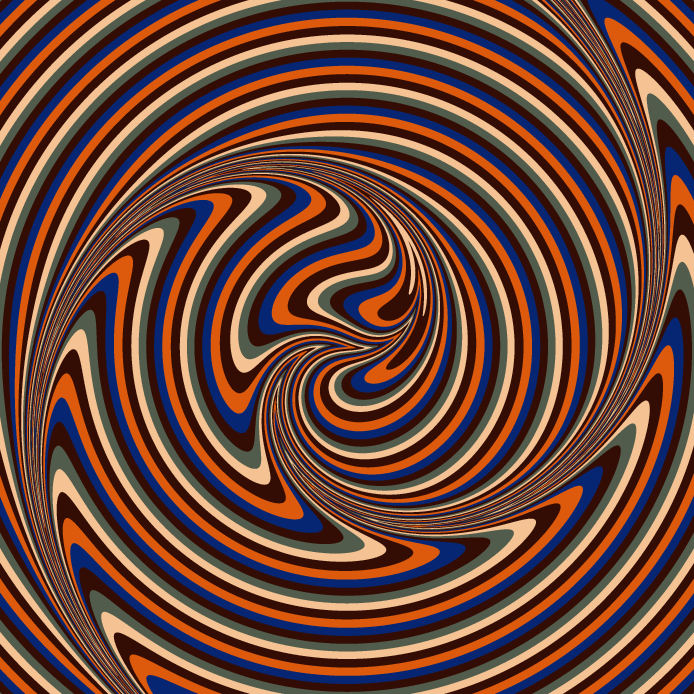}{200pt}\hfill&
\hfill\figscale{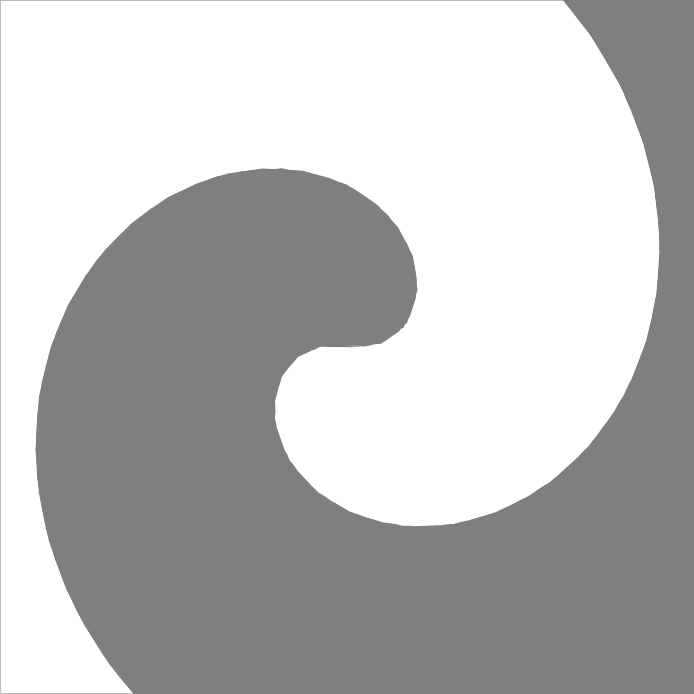}{200pt}\hfill&
\cr\+
\hfill\figdef{vortexline}\hfill&
\hfill\figdef{yinyang}\hfill&
\cr}

\section{French Curl}

\vbox{\settabs 2\columns\+
\hfill\figscale{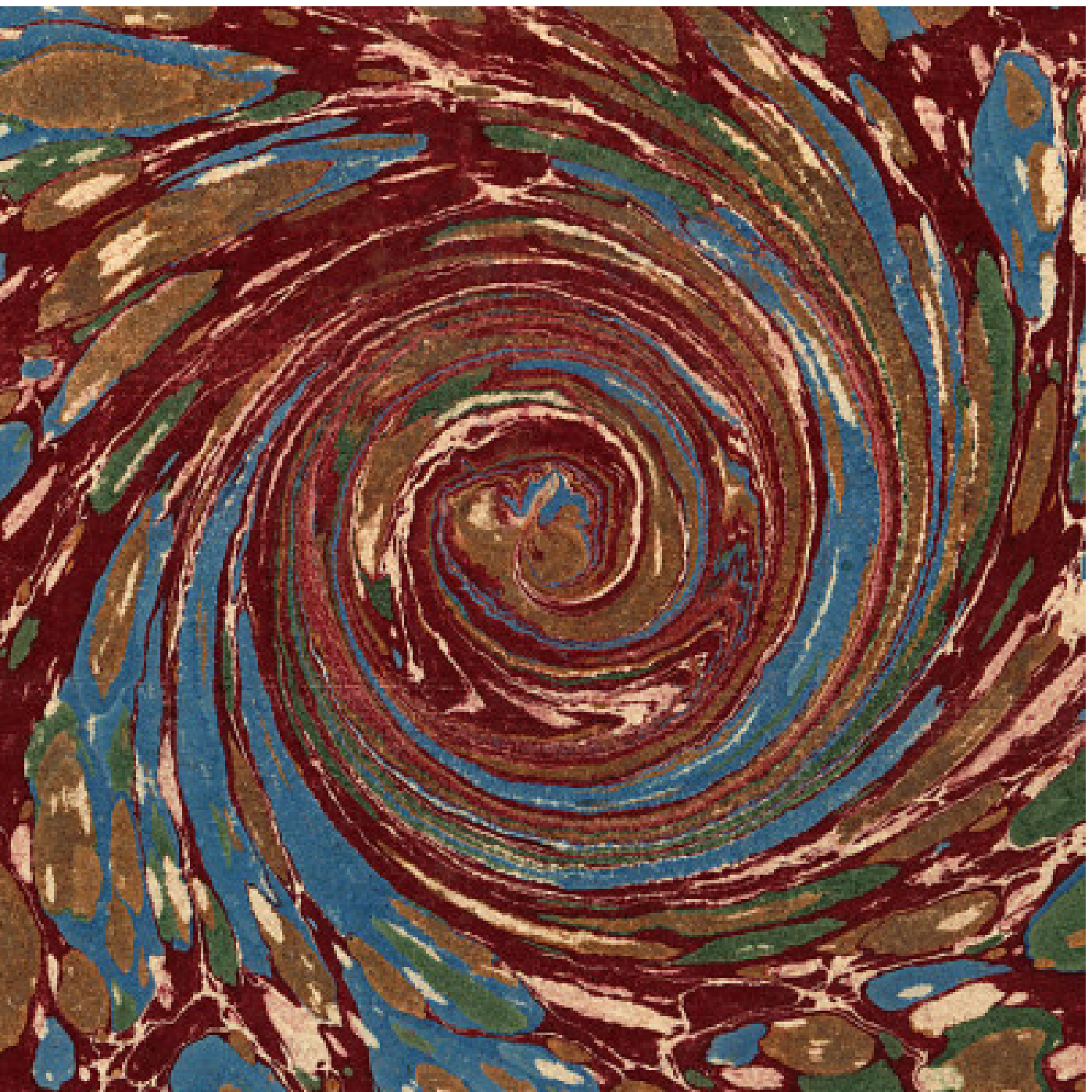}{200pt}\hfill&
\hfill\figscale{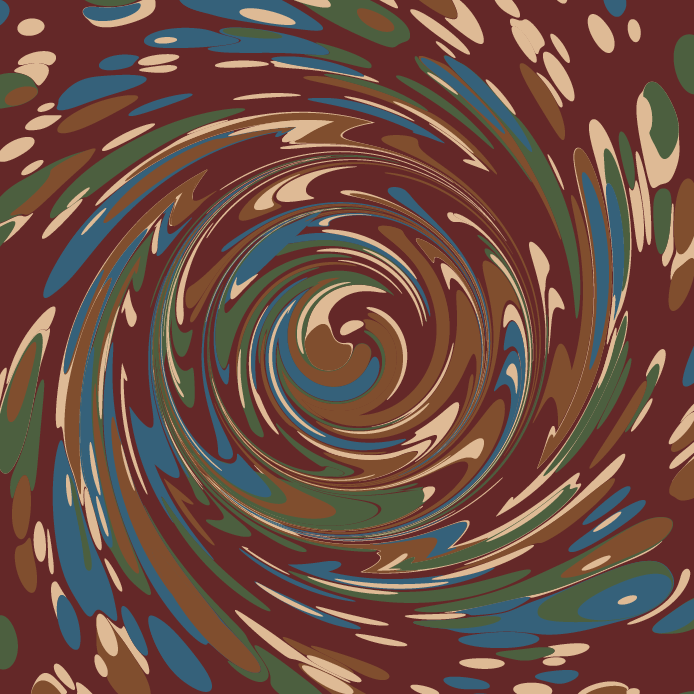}{200pt}\hfill&
\cr\+
\hfill\figdef{FrenchCurl}\hfill&
\hfill\figdef{curl2}\hfill&
\cr}

 At first glance the photographic detail of an 1880 French Curl
 marbling\numberedfootnote{Creator of the paper unknown. Scanned by
 Aristeas from a book in his own possession. [Public domain], from
 Wikimedia Commons} in \figref{FrenchCurl} looks as though it
 was produced by a vortex.  The white nested chevrons show that it is
 instead the product of concentric circling by a
 stylus.  \figref{curl2} is a mathematically produced marbling of a
 stylus circling four times at various radii; it has chevrons and much
 of the character of the original.

\vbox{\settabs 2\columns\+
\hfill\figscale{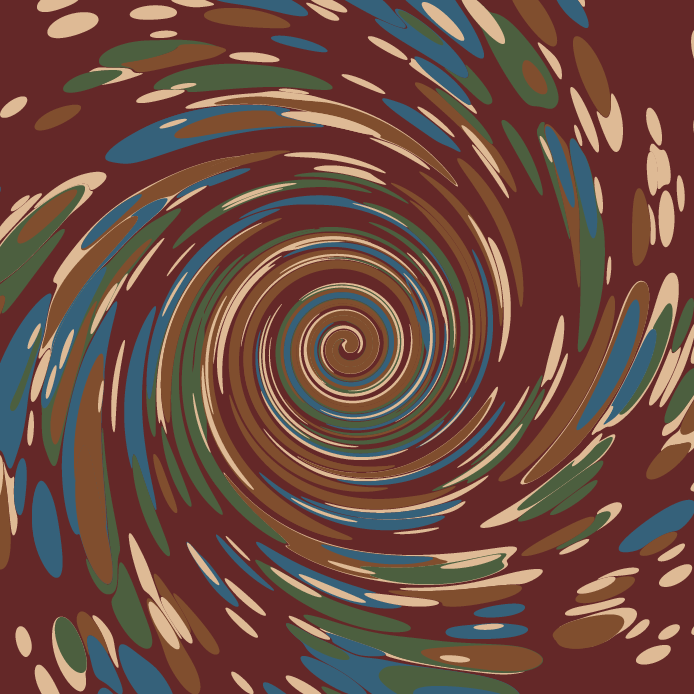}{200pt}\hfill&
\hfill\figscale{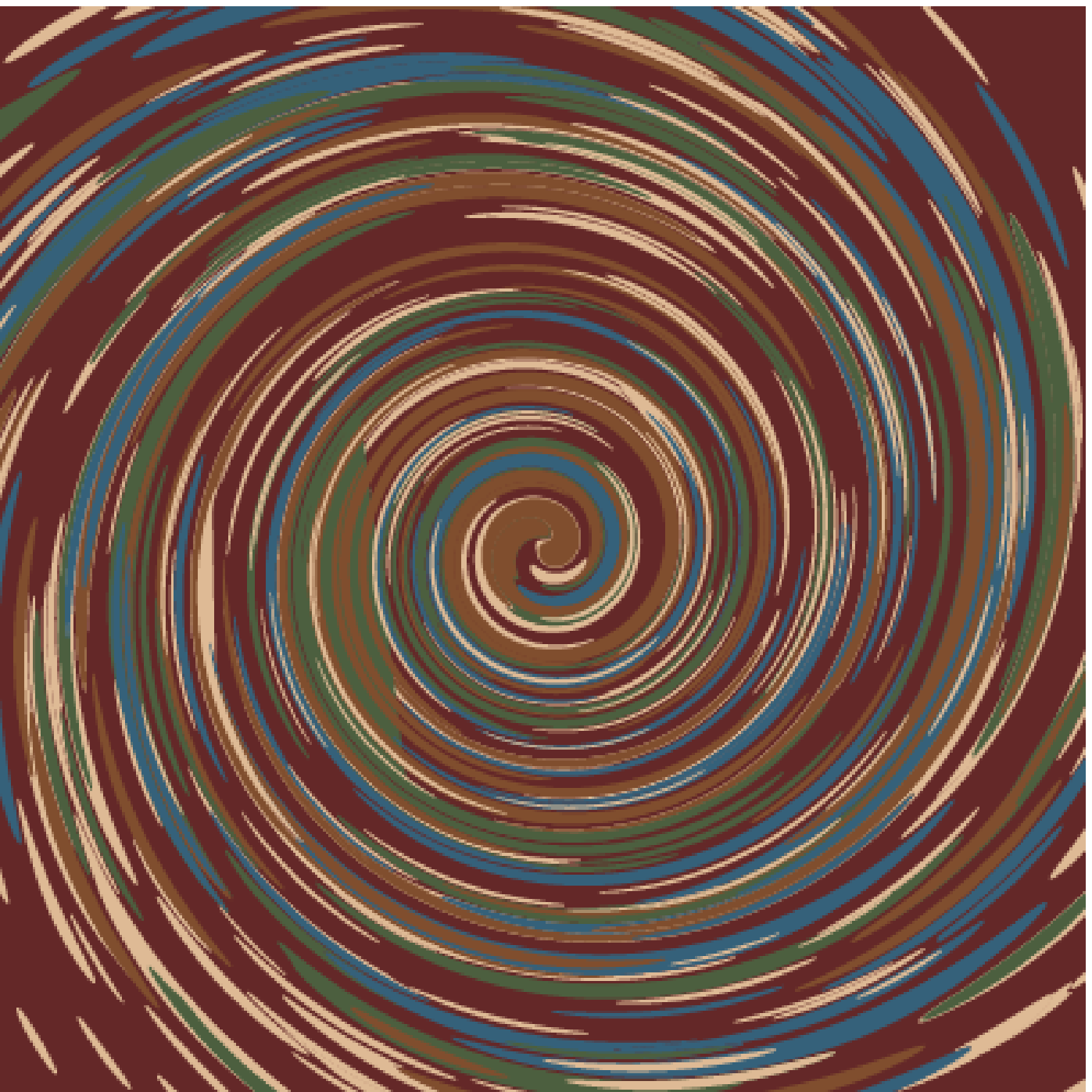}{200pt}\hfill&
\cr\+
\hfill\figdef{curlamboseen}\hfill&
\hfill\figdef{curlamboseen5}\hfill&
\cr}

 A Lamb-Oseen vortex captured immediately after the stirring at its
 center produces \figref{curlamboseen}, which bears some likeness
 to \figref{FrenchCurl}.  And this would be captured if the
 paper could be laid into and peeled off the tank quickly after
 stirring the vortex.

 But within a minute, the rotational shear has propagated outward as
 shown in \figref{curlamboseen5}, which no longer
 resembles \figref{FrenchCurl}.

\section{Bubbles}

 One way to produce vortexes in a Newtonian fluid is to move a stylus
 briskly in a straight line.  If the Reynolds number exceeds 90, then
 it will shed Karman vortexes to alternating sides of the stylus path.

 A 25~mm diameter dowel submerged 12.5~mm has a characteristic-length
 $D=0.0042$m, the same as a half-submerged 25~mm diameter sphere.  If
 the kinematic viscosity ($\nu=0.001~{\rm m^2/s}$) of the liquid is
 1000 times that of water and the dowel is moved 20~cm/s through the
 liquid, then the Reynolds number is about 0.85, far less than the 90
 needed to spawn vortexes.  Re is inversely proportional to viscosity;
 reducing the viscosity by a factor of 10 raises Re to 8.5.

 A half sphere of diameter 25~mm has buoyant-pressure (restoring force
 divided by cross-section area) of about $81~\rm N/m^2$.  Surface
 tension pressure (restoring force divided by cross-section area) of
 water is roughly $3.7~{\rm N/m^2}$.

 Drag is the force on the object moving through the tank fluid.  There
 must be an equal and opposite net force on the liquid.  Drag $D$ for
 a sphere is the product of the friction coefficient $C_D$, frontal
 area ($\pi\,d^2/4$), and dynamic head $V^2\rho/2$ (for water
 $\rho=997\rm~kg/m^3$).  That force divided by the frontal area of the
 object is a pressure (suction actually).

 A bubble will be formed if this suction behind the moving stylus is
 larger than the sum of the restoring forces at the liquid surface.

 For kinematic viscosity $\nu=1000\rm~mm^2/s$ (1000 times that of
 water) the suction behind the 25~mm diameter dowel or sphere is
 $88\rm~N/m^2$, which exceeds the restoring pressures $81\rm~N/m^2$
 and $3.7\rm~N/m^2$, and bubbles can result.

 Kinematic viscosities below $50\rm~mm^2/s$ (50 times that of water)
 would be needed to spawn vortexes in marbling.  At $50\rm~mm^2/s$
 viscosity, a 25~mm cylinder would need to be moved at 20~cm/s over at
 least 16~cm before the first vortex was shed.

 In water, a 5~mm cylinder moving at 2~cm/s would shed vortexes 3~cm
 apart.  A 1~mm diameter stylus moved in a straight path at 5~cm/s
 would not shed vortexes.

 So existing evidence of Karman (shed) vortexes is only likely to be
 found in marbling produced on a tank filled with water.

\beginsection{References}

\bibliographystyle{unsrt}
\bibliography{vortex}

\vfill\eject
\bye